\documentclass[useAMS,prd,aps,floats,twocolumn]{revtex4}
\usepackage{pslatex}
\usepackage{graphicx}
\def\nn{\nonumber}

\def\refe@jnl#1{{#1}}
\def\aj{\refe@jnl{Astron.~J.}}
\def\araa{\refe@jnl{Annu.~Rev.~Astron.~Astrophys.}}
\def\apj{\refe@jnl{Astrophys.~J.}}
\def\apjl{\refe@jnl{Astrophys.~J.~Lett.}}
\def\aap{\refe@jnl{Astron.~Astrophys.}}
\def\mnras{\refe@jnl{Mon.~Not.~R.~Astron.~Soc.}}
\def\prd{\refe@jnl{Phys.~Rev.~D}}
\def\fcp{\refe@jnl{Fund.~Cos.~Phys.}}
\def\physrep{\refe@jnl{Phys.~Rep.}}
\def\physlett{\refe@jnl{Phys.~Lett.}}

\def\DiracSlash#1{\slash{ \hspace{-0.2cm}#1}}
\def\DiracMatrix#1{\gamma_{#1}}

\def\ar{{c_r}}
\def\al{{c_l}}
\def\xm{{x_-}}
\def\xp{{x_+}}
\def\xpm{{x_\pm}}
\def\dm{{\rm dm}}
\def\dms{{{\rm dm}^\ast}}
\def\pdm{{p_\dm}}
\def\pdms{{p_\dms}}
\def\pe{{p_{e^-}}}
\def\pp{{p_{e^+}}}
\def\pg{{p_{\gamma}}}
\def\xb{{\bar x}}
\def\xg{{x_\gamma}}
\def\me{{m_e}}
\def\MF{{m_F}}
\def\mdm{{m_\dm}}

\def\PL{{P_l}}
\def\PR{{P_R}}

\begin{document}

\title{Revisiting Bremsstrahlung emission associated with
  Light Dark Matter annihilations}

\author{C. Boehm\footnote{On leave from LAPTH, UMR 5108, 9 chemin de Bellevue - BP 110,
  74941 Annecy-Le-Vieux, France.} and P. Uwer\footnote[6]{Heisenberg fellow of the Deutsche
Forschungsgemeinschaft.}}

\affiliation{Physics Department, Theory Unit, CERN, CH-1211 Geneva 23,  Switzerland}

%\date{June 2, 2006}

\begin{abstract}
  We compute the single bremsstrahlung emission associated with the
  pair annihilation of spin-0 particles into electrons and positrons, via
  the $t$-channel exchange of a heavy fermion. We compare our result
  with the work of Beacom
  et al. \cite{beacom05}.
  Unlike what is stated in the literature, we show that the Bremsstrahlung cross section is not necessarily given by the
  tree-level annihilation cross section (for a generalized kinematics) times
  a factor related to the emission of a soft photon. Such a factorization
  appears only in the soft photon limit or in the approximation where the masses of the particles
  in the initial and final states are negligible with respect to the mass of
  the internal particle.
  However, in the latter case, we do not recover the same
  factor as for $e^+ e^- \rightarrow \mu^+ \mu^- \gamma$.
  Numerically the difference, in the hard photon limit, is as large as a factor 3.6.
  However the effect on the upper limit of the dark matter mass is not significant.
  Using gamma ray observations, we obtain an upper limit on the dark matter
  mass of 30 or 100 MeV (depending on the region that is considered for the 511 keV analysis)
  while Beacom et al. found 20 MeV.
\end{abstract}
\maketitle

%%%%%%%%%%%%%%%%%%%%%%%%%%%%%%%%%%%%%%%%%%%%%%%%%%%%%%%%%%%%%%%%%%%%%%
\section{Introduction and set up}
%%%%%%%%%%%%%%%%%%%%%%%%%%%%%%%%%%%%%%%%%%%%%%%%%%%%%%%%%%%%%%%%%%%%%%
The Milky Way radiation has been mapped extensively at different
wavelengths for many decades. Photon emission at low and ultra-high
energy is still currently studied. All the latest observations
appear extremely exciting and shed new light on the physics of the
galactic centre. Among them, there is the observation of a 511 keV
emission line by SPI, a spectrometer on board of the INTEGRAL
satellite launched in 2002. The 511~keV line has been detected by
many experiments since the seventies (see
Ref.~\cite{Johnson72,Leventhal78,Kinzer01}), but the origin
of the line could not be established unambiguously. With the new
results from SPI \cite{Jean03,Jean:2005af,Knodlseder05}, 
one could determine that it is due to
(para-)positronium formation, thus confirming the presence of anti
matter in our galaxy.

A very striking feature of this line is that it is mostly emitted in
the bulge of the Milky Way (the distribution is well approximated by
a sphere; the full width at half maximum is about $10^\circ$ of
diameter). No significant emission
was detected in the disk. This characteristic tends to exclude
astrophysical sources such as hypernovae/Wolf--Rayet stars, pulsars,
and cosmic ray interactions. On the other hand, this observation
could point towards an old galactic population
\cite{weidenspointner}.

A blind source analysis showed that more than 8 point sources could
explain the emission \cite{Knodlseder05}.  A diffuse source provides
a very good fit too. None of these sources has been found as yet.
This makes the origin of the galactic positrons even more
mysterious.

Low Mass X-ray Binaries (LMXB) could perhaps be an explanation if
the positrons emitted in the disk escape into the bulge. Their
characteristics are poorly known though (e.g. flux, jet content). No
511 keV emission has been seen so far from the brightest of these
objects, but more observations will be dedicated to this kind of
candidates. Type 1a supernovae (SN1a) may also explain the line
emission if the positron escape fraction and the supernovae
explosion rate are large enough to maintain a steady flux. At
present various observations contradict with each other and it is
quite difficult to know whether these two conditions are indeed
satisfied in our galaxy \footnote{For example, their contribution to
  $\gamma$ ray background may have been overestimated \cite{Ahn:2005ws}.}. 
Dedicated observations of SN1a remnants are
also scheduled with INTEGRAL and should help answering these
questions \cite{weidenspointner,weidenspointnerbis}.

Another possible source of low energy positrons could be dark matter annihilations \cite{511}. This scenario
requires a dark matter mass between a few MeV and a few hundred MeV, two kinds of interactions, and a cuspy dark
matter halo profile (likely to be as cuspy as a Navarro--Frenk--White model \cite{NFW}). This dark matter profile
corresponds to theoretical predictions from numerical simulations, but it has not been confirmed as yet
observationally \cite{be,helmi02}.

Light dark matter (LDM) particles were initially proposed in the
context of the so-called cold dark matter crisis. These particles
were meant to experience a new damping effect, including one that is
half-way between the collisional damping and the free-streaming.
They were meant to inherit from the damping of relativistic species
($X = {\gamma, \nu}$) with which they weakly interact either
directly or indirectly \cite{bfs,boehmthesis,brhs,boehmschaeffer,boehmS}.
In the indirect case, the damping of the species $X$ is communicated
to dark matter through the dark matter coupling to other particles
in contact with $X$. For example, before recombination,  dark matter
indirectly experiences the Silk damping, until it thermally
decouples from electrons. This new effect (called induced damping)
is simply the generalization of the Silk damping effect to dark
matter and has been confirmed numerically \cite{loeb}. In the direct
case, LDM may suffer from the mixed damping effect
\cite{bfs,boehmthesis,boehmschaeffer}, which corresponds to the
situation where, due to a weak but large enough neutrino--DM
interaction rate, dark matter inherits from the neutrino
free-streaming. This occurs when LDM stays coupled to neutrinos
after their thermal decoupling. For some specific values of the
DM--$\nu$ elastic scattering cross section, these interactions
affect the DM fluid but do not change the neutrino fluid properties.
This effect was stressed in Refs.~\cite{bfs,boehmthesis} for
neutralinos when they are strongly mass degenerated with sneutrinos
and for MeV particles having weak interactions
\cite{bfs,boehmthesis,boehmS}.

Light dark matter particles therefore have this interesting
property: on the one hand, they look like cold dark matter
candidates because of their mass and weak interactions. On the other
hand, their linear matter power spectrum is cut--off at a 
cosmological scale (due to these new damping effects), which is a
well-known characteristic of warm dark matter candidates.

To avoid an overproduction of gamma--rays and satisfy simultaneously
the relic density criterion, the LDM total annihilation cross
section at early times should be larger than that in the primordial
universe or in virialized objects \cite{bens}. This leads to a
scenario in which the LDM interactions, and hence the damping, are
suppressed \cite{boehmS} (the damping mass in the linear matter
power spectrum was estimated to be about $100 \ M_{\odot}$).

Following up Ref.~\cite{511}, there have been many other dark matter
models proposed to explain the 511 keV emission, e.g.
Refs.~\cite{Picciotto:2004rp,Hooper:2004qf,Bertone:2004ek,%
Oaknin:2004mn,Ferrer:2005xv,Gunion:2005rw}. Here we  focus 
on the annihilating case. In this scenario, there is a light extra
gauge boson $Z'$ (needed for the relic density) and
heavy charged particles (for the 511 keV emission) \cite{bens}. 
This new set of
particles is quite similar to that expected in $N=2$ supersymmetry.

The model is described in detail in Ref.~\cite{bf}. The constraints on the couplings are given in
Ref.~\cite{boehm,boehmascasibar,ascasibar}. We disregard the possibility that dark matter is fermionic
\footnote{The mass range of the exchanged charged particle has been already excluded by LEP data in the case of a
velocity-independent cross section while the case of a pure velocity-dependent cross section would yield a very
unlikely DM halo profile. Note, however, that in principle it is premature to exclude spin$-1/2$ particles before
a careful treatment of the positron propagation has been performed. Nevertheless, with a profile at least as cuspy
as in Ref.~\cite{ascasibar}, the emission is naturally confined in the bulge, so the propagation can be neglected
and the conclusion should remain the same.}, following up Ref.~\cite{ascasibar}.

Heavy dark matter particles, i.e. $100 \ \mbox{MeV}>\mdm > 10$ MeV,
are in danger of yielding a too large contribution to the anomalous
magnetic moment of the electron \cite{boehmascasibar,ascasibar}.
Also they would make the observed fraction of para and
ortho-positronium very hard to explain \cite{Jean:2005af} and would
eventually yield a too large $e^+ e^-$ inflight annihilation
contribution \cite{beacom06}.

Any other way of reducing the mass range which is not too model dependent and which could be related to the 511
keV flux is very interesting. In that context, the Beacom et al. proposal \cite{beacom05}, to constrain the dark
matter mass from the comparison between observations and the gamma--ray flux originating from the Bremsstrahlung,
looks promising even if it gives a smaller contribution than the inflight annihilations.

This suggestion is complementary to a previous estimate made
before INTEGRAL's publication \cite{bens}. In that work, it was
assumed that each electron and positron produced by dark matter
annihilations spontaneously generates one photon after their
emission. The authors assumed that the total annihilation cross
section was either velocity-dependent or velocity-independent. After
comparison with the  gamma--ray observations (and requiring the
correct relic density), they concluded that only a
velocity-dependent cross section was possible for $\mdm < 100$ MeV
while for greater masses either a velocity-dependent or independent
cross section could be allowed.

However, the model that fits INTEGRAL/SPI data requires two kinds of
interactions simultaneously. 
One is velocity-dependent and is needed for fitting
the relic density. The other one is velocity-independent and is
quite mandatory for fitting the 511 keV emission morphology. Such a
scheme is not compatible with Ref.~\cite{bens} in which it was
assumed that only one process (either velocity dependent or
independent) was contributing to the total annihilation cross
section. One therefore has to estimate the gamma ray production
again  and derive the corresponding upper bound on the dark matter
mass, assuming that LDM is at the origin of all the 511 keV emission
mapped by SPI.

By doing so the authors in Ref.~\cite{beacom05} concluded that $\mdm$ could not exceed 20 MeV. Their estimate of
$\sigma_{\dm \ \dm^\ast \rightarrow e^+ e^- \gamma}$ is based on an old observation made by Berends et al.
\cite{Bberendsetalb} according to which, in QED or $SU(N)$ gauge theories, the cross sections associated with the
single emission of a hard bremsstrahlung photon becomes remarkably simple in the ultrarelativistic limit. Using
specific examples, Berends et al. showed that a factorization into two terms can be observed. One factor is the
squared amplitude associated with the 2-to-2 process (i.e. without the additional emission of the photon), and
evaluated for a generalized kinematics. The second factor is related to the infrared factor, which describes the
soft photon limit. Such a property shows up for example in $e^+ e^- \rightarrow \mu^+ \mu^- \gamma$. The
expressions for the initial and final-state radiation cross sections associated with this process can be found for
example in the review of Berends and B\"ohm \cite{berendsbohm}.

Beacom et al. thus argued that the final-state radiation cross
section of dark matter particles annihilating while they are almost
at rest can be written as the product of the two-to-two process
times an universal factor. In a next step they assumed that this
factor can be obtained from the known results for the final-state
radiation associated with the $e^+ e^- \rightarrow \mu^+ \mu^-
\gamma$ process in the ultrarelativistic limit, where the electron
mass is neglected.

However, in the case of dark matter, the ultrarelativistic limit is not appropriate. Furthermore for chiral dark
matter couplings it is also obvious that the electron mass cannot be neglected. Also $e^+ e^- \rightarrow \mu^+
\mu^-$ and $\dm \ \dm^\ast \rightarrow e^+ e^- $ are intrinsically different.  $e^+ e^- \rightarrow \mu^+ \mu^-$
is due to  a photon exchange in the $s$-channel; it involves only vectorial couplings. In contrast, $\dm \
\dm^\ast \rightarrow e^+ e^- $ proceeds through a $t$-channel exchange
of a heavy fermion and can involve chirality flipping couplings.
Thus writing the Bremsstrahlung cross section as the product of two factors may not be correct.

Here we demonstrate that the approach followed in Ref.~\cite{beacom05} fails to correctly describe the
Bremsstrahlung process in the hard photon limit.

%%%%%%%%%%%%%%%%%%%%%%%%%%%%%%%%%%%%%%%%%%%%%%%%%%%%%%%%%%%%%%%%%%%%%%
\section{Annihilation and Bremsstrahlung cross section}
%%%%%%%%%%%%%%%%%%%%%%%%%%%%%%%%%%%%%%%%%%%%%%%%%%%%%%%%%%%%%%%%%%%%%%
\begin{figure}[htbp]
  \begin{center}
    \leavevmode
    \includegraphics[width=8.cm]{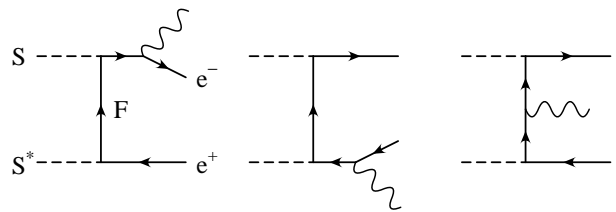}
    \caption{Feynman diagrams contribution to $S+S^\ast\to e^++e^-+\gamma$.}
    \label{fig:FeynmanDiagrams}
  \end{center}
\end{figure}
In what follows we study the process
\begin{equation}
  S(\pdm) + S^\ast(\pdms) \to e^-(\pe) + e^+(\pp) + \gamma(\pg),
\end{equation}
where $S$ denotes a spin-0 dark matter particle.

Since dark matter is neutral, the emission of a photon necessarily
comes from the final state particles and the particle that is
exchanged during the annihilation process ($F$). The corresponding
Feynman diagrams are shown in Fig. \ref{fig:FeynmanDiagrams}. It is
convenient to use the scaled energies as independent variables:
\begin{equation}
  x={2 k\cdot \pe\over s}, \quad
  \xb={2 k\cdot \pp\over s}, \quad
  \xg={2 k\cdot \pg\over s},
\end{equation}
where $k=\pdm+\pdms = \pe+\pp+\pg$ and $s=(\pdm+\pdms)^2$. Due to
energy momentum conservation, $x, \xb,$ and  $\xg$ are not
independent of each other:
\begin{equation}
  2 = x + \xb + \xg.
\end{equation}
In terms of  $x, \xb$ and $\xg$, the scalar products between final
state momenta read:
\begin{eqnarray}
  \pp\cdot \pg  &=& {s\over 2} \* (1 - x) ,  \\
  \pe\cdot\pg &=& {s\over 2} (1 - \bar{x}) ,\\
  \pe\cdot\pp &=& {s\over 2} (1- \xg - 2 z),
\end{eqnarray}
where $z= \me^2/s$ and $\me$  is the electron mass. In the case of
dark matter it is legitimate to assume that it is almost at rest
when it annihilates. At Earth position, the dark matter velocity
dispersion is about three orders of magnitude smaller than the speed
of light ($v \sim 10^{-3} c$). It is even smaller below 1 kpc (say,
inside the galactic centre). Thus any scalar product that involves
the four-momenta of both initial and final state particles can be
expressed in a very simple way, without involving any undefined
angles. In particular we obtain:
\begin{eqnarray}
  \pdm\cdot \pp &\approx&  {s\over 4} \xb, \\
  \pdm\cdot \pe &\approx& {s\over 4} x,\\
  \pg\cdot \pdm &\approx& {s\over 4} \xg.
\end{eqnarray}

\subsection{The annihilation cross section}
For later use we reproduce here the annihilation cross section
associated with the process
\begin{equation}
   S(\pdm) + S^\ast(\pdms) \to e^-(\pe) + e^+(\pp)
\end{equation}
at tree--level \cite{bf}. Neglecting the P-wave annihilation cross
section because of the suppression of the dark matter velocity in
the 511 keV region, we find that the corresponding cross section
$\sigma_0$ is given by:
\begin{eqnarray}
  \sigma_0
  &=&
  {1\over 16\*\pi}\*{1\over s}\* {\beta_e \over \beta_i }
  |M_0|^2\nn\\
  &=&
  {1\over 32\*\pi}\*{\beta_e^3 \over \beta_i }
    \*{(2\*\ar\*\al \* \MF +
   \me\*(\al^2+\ar^2))^2
   \over (\MF^2+\mdm^2-\me^2)^2},
\label{eq:sigma0}
\end{eqnarray}
with
\begin{equation}
  \beta_e = \sqrt{1-4z}
\end{equation}
and
\begin{equation}
  \beta_i = \sqrt{1-{4\mdm^2\over s}},
\end{equation}
where $\mdm$ denotes the mass of the annihilating scalar and $\MF$
the mass of the heavy charged fermion that is exchanged in  the
$t$-channel. From inspection of Eq.~(\ref{eq:sigma0}), one can
easily see that in general it is not appropriate to neglect the
electron mass. If the ratio between the chiral couplings $\al, \ar$
is of the order of $\me/\MF$  then the term that is formally suppressed 
through the electron mass
has the same importance as the term in $m_F$.  
Such a scenario arises naturally in 2-Higgs doublet models with
flavour changing neutral couplings. I.e. the lagrangian
\begin{equation}
  {\cal L } = - g_l {\me \over v} \bar e_R F_L S 
  - g_r {\MF \over v} \overline{F}_R e_L S + h.c. 
\end{equation}
leads immediately to ${\al\over \ar} = {\me\over \MF}$ if we identify
$\al = g_l {\me \over v}$ and $\ar = g_r {\MF \over v}$.
Such a case does not
fit SPI data unless one considers extremely large couplings of a few
units. Since the particle $F$ is charged, the mass $\MF$ must be
larger than 100 GeV, given LEP limits \cite{LEP} restricting the range
of the couplings that can be considered.

In the limit where $\mdm$ and $m_e \ll \MF$, the cross section becomes:
\begin{equation}
   \sigma_0 = {1\over 32\*\pi}\*{\beta_e^3 \over \beta_i }
  \*{1\over \MF^2} \*\kappa(\al,\ar,\me,\mdm,\MF)+O\left({1\over \MF^5}\right),
\end{equation}
with
\begin{eqnarray}
  \kappa(\al,\ar,\me,\mdm,\MF) &=&
  \left(2\*\al\*\ar+(\al^2+\ar^2)\*{\me\over \MF}\right)^2 \nn \\
  &+& 8 \ar^2\*\al^2\*{\me^2-\mdm^2\over \MF^2}.
\end{eqnarray}
Keeping only the leading term in the $1/\MF$  expansion, we obtain for $\sigma_0$
\begin{equation}
   {\al^2\ar^2  \*\beta_e^3\over 8\*\pi\beta_i\MF^2}  +O\left({1\over
   \MF^3}\right).
\end{equation}

To compute the annihilation rate and therefore the flux of positrons
and/or photons that are emitted during the dark matter
annihilations, one must consider $\sigma_0 v_r$ (where $v_r$ is the
dark matter relative velocity) instead of $\sigma_0$. $v_r$ is twice
the dark matter velocity in the centre-of-mass frame (written here
as $\beta_i$). We therefore recover the results of Ref.~\cite{bf}.

\subsection{Bremsstrahlung cross section}
\subsubsection{General formulae}
In terms of the squared matrix element $M_\gamma$, the single
differential bremsstrahlung cross section reads
\begin{equation}
\frac{d \sigma_\gamma}{d \xg} =
   { 1\over  256 \pi^3  \beta_i }
   \int_{\xm}^{\xp}  dx \ |M_\gamma|^2 \nn,
\end{equation}
where the bounds $\xm, \xp$ are given by
\begin{eqnarray}
  \xpm &=&  {1\over 2}\*
  \left(2-\xg
  \pm \xg\*\sqrt{1- {4\*z\over 1-\xg}}\right).
\end{eqnarray}
The squared matrix element is obtained from the evaluation of the Feynman diagrams shown in Fig.
\ref{fig:FeynmanDiagrams}. Since there are many different mass scales, the complete result -- keeping the full
mass dependence -- is rather lengthy. We find:
\begin{eqnarray}
\label{wholeexpression}
  |M_\gamma|^2 &=& {- e^2 \over {\cal E} } \*
  \bigg\{ \MF^4 \* {\cal A } \* \bigg[ \Big( 2\* \al \* \ar
  \* m_F  + (\al^2 + \ar^2) \* \me \Big)^2
  \nn\\
  && \hspace{1.cm} - 8\* \al^2 \* \ar^2 \* \Big((2\*\xg - 1)\*\mdm^2 +
  \me^2 \Big) \bigg]
  \nn \\
  &+&   8 \* \al \* \ar \* (\al^2 + \ar^2) \* \me \* \MF^3 \* {\cal B}
  \nn\\
  &+& 2 \* \MF^2 \*  \Big(\me^2  (\al^4+ \ar^4)\*  {\cal C} +
  2\* cl^2\* cr^2 \* \mdm^2\* {\cal D} \Big)
  \nn\\
  &-& 4 \* \al \* \ar \* (\al^2 + \ar^2) \* \me\*\MF
   \Big(\me^2 + \mdm^2 \* (2\* \xg-1) \Big) \* {\cal C}
   \nn \\
  &+&  {\cal N }
  \bigg\},
\end{eqnarray}
where $e$ denotes the electric charge and where the explicit results for the functions $\cal A, B, C, D, E, N$ are
given in the appendix. We note that in addition to the symmetry in $\al,\ar$, there is also a symmetry in $x
\leftrightarrow \xb$, when using $\xb$ instead of $\xg$. Furthermore we have checked that we reproduce the correct
factorization when the emitted photon becomes soft:
\begin{eqnarray}
  &&|M_\gamma|^2 \ \ \stackrel{\xg \, \to \, 0} {\longrightarrow}
  \nn\\
  &&e^2\left( {2(\pe\cdot\pp)\over (\pe\cdot\pg)\* (\pp\cdot\pg)}
    - { \me^2\over (\pe\cdot\pg)^2}
    -  {\me^2\over (\pp\cdot\pg)^2}
    \right) |M_0|^2. \nn\\
\end{eqnarray}
For arbitrary hard photons no such factorization can in general be
observed.

There are many extra terms in the squared amplitude with respect to that of the annihilation. Some are
particularly relevant when $m_F/2 \lesssim \mdm \lesssim m_F$ (e.g. $\propto \xg \, {\cal{A}}$, ${\cal{D}}$,
${\cal{N}}$) or in the case of very large couplings, i.e. $(c_l^2 + c_r^2) \ m_e \ \gtrsim \ 2 \, c_l \, c_r \,
m_F$.

Hence from Eq.~(\ref{wholeexpression}) one readily sees that -- unless one makes simplifying assumptions -- the
cross section for the emission of an additional photon cannot be
written as 
the tree-level cross section for a generalized kinematics times a
factor related to the soft photon emission.

The integration of the squared matrix element $|M_\gamma|^2$ over $x$ to obtain the photon spectrum is
straightforward. In the following we only present analytic results in the approximation that the dark matter mass
is small with respect to the mass of the particle $F$.

\subsubsection{Results in the limit  $\MF \gg ( \mdm, \me )$}

Expanding Eq.~\ref{wholeexpression} in terms of $1/\MF$ and keeping
the dominant terms up to $1/m_F^4$, we obtain:
\begin{eqnarray}
  |M_\gamma|^2 &=&{e^2 \ \* \ \kappa(\al,\ar,s,\me,\MF)\over
    s^2\*\MF^2}\nn\\
  &&\hspace{-1.2cm}\times\bigg\{
  8\*\me^2\*(\me^2-\mdm^2)
  \left( {1\over (1-x-\xg)^2}+{1\over (1-x)^2}\right) \nn\\
  &&\hspace{-1.2cm}-{16\over \xg}\*\Big(\me^4-(3-2\*\xg)\*\mdm^2\*\me^2
  +(\xg^2-2\*\xg+2)\*\mdm^4\Big)\nn\\
  &&\times\left( {1\over 1-x-\xg} -{1\over 1-x} \right)
\bigg\} + O\left(\frac{1}{\MF^5}\right).
\end{eqnarray}
Integrating over $x$ we obtain the photon spectrum:
\begin{eqnarray}
   {d\sigma_\gamma\over d\xg} &=& \ \sigma_0 \
   { \alpha \over   \pi} \ {1 \over \beta_e^3} \  {1 \over s^2}
   \nn\\
  &&\hspace{-1.2cm}\times\bigg\{
  2\*s\*(-s+4\*\me^2)
  {1\over \xg}\*\sqrt{(1-4\*z-\xg)\*(1-\xg)}\nn\\
  &&\hspace{-1.2cm}+{16\over \xg}\*\Big(\me^4-(3-2\*\xg)\*\mdm^2\*\me^2
  +(\xg^2-2\*\xg+2)\*\mdm^4\Big)\nn\\
 && \times\ln\left( {1-\xm\over 1-\xp} \right) \bigg\} + O\left(\frac{1}{\MF^5}\right).
\end{eqnarray}
Keeping the electron mass only in the logarithm so as to regularize the
mass singularity, the above result can be
simplified further:
\begin{eqnarray}
    {d\sigma_\gamma\over d\xg} &\approx&
  \sigma_0 {\alpha\over \pi}
 \*{1\over \xg}\* \bigg\{
 \bigg(1+{s'^2\over s^2}\bigg)
 \ln \left( {s'\over \me^2} \right)   - 2  \*{s'\over s}
 \bigg\},
  \label{eq:ScalarFSR}
\end{eqnarray}
where we introduced
\begin{eqnarray}
  s' = (\pe+\pp)^2 = s \  (1-\xg).
\end{eqnarray}

For  $ e^+ \, + \, e^- \, \to \, \mu^+ \, + \, \mu^- + \, \gamma$, keeping only the final-state radiation,
 the formula  is given by \cite{berendsbohm}:
\begin{eqnarray}
  &&{d\sigma_{e^+ + e^- \to \mu^+ + \mu^- + \gamma}\over d\xg} \ = \  \sigma_{e^+ + e^- \to \mu^+ + \mu^-} \nn\\
  &&
  \times {\alpha\over \pi}
  \*{1\over \xg}\*
  \left(1+{s'^2\over s^2}\right)
 \left(\ln \left( {s'\over m_\mu^2} \right)  - 1\right)
  \label{eq:MuonFSR}
\end{eqnarray}

Comparing Eq.~(\ref{eq:ScalarFSR}) and Eq.~(\ref{eq:MuonFSR}), it is
 easy to see
that the bremsstrahlung cross section associated with the
annihilation of two scalars cannot be obtained from the final-state
radiation in muon pair production.  However, in the soft photon
limit ($s' \rightarrow s$) the two results agree after the
identification $m_\mu \to \me$, as it must be. We are thus led to
the conclusion that the bremsstrahlung correction to the
annihilation of two scalars cannot be evaluated as it was done in
Ref.~\cite{beacom05}.

%%%%%%%%%%%%%%%%%%%%%%%%%%%%%%%%%%%%%%%%%%%%%%%%%%%%%%%%%%%%%%%%%%%%%%
\section{Discussion and upper limit on the dark matter mass}
%%%%%%%%%%%%%%%%%%%%%%%%%%%%%%%%%%%%%%%%%%%%%%%%%%%%%%%%%%%%%%%%%%%%%%

\begin{figure}[htpb]
\begin{center}
\includegraphics[angle=-90,width=11.cm]{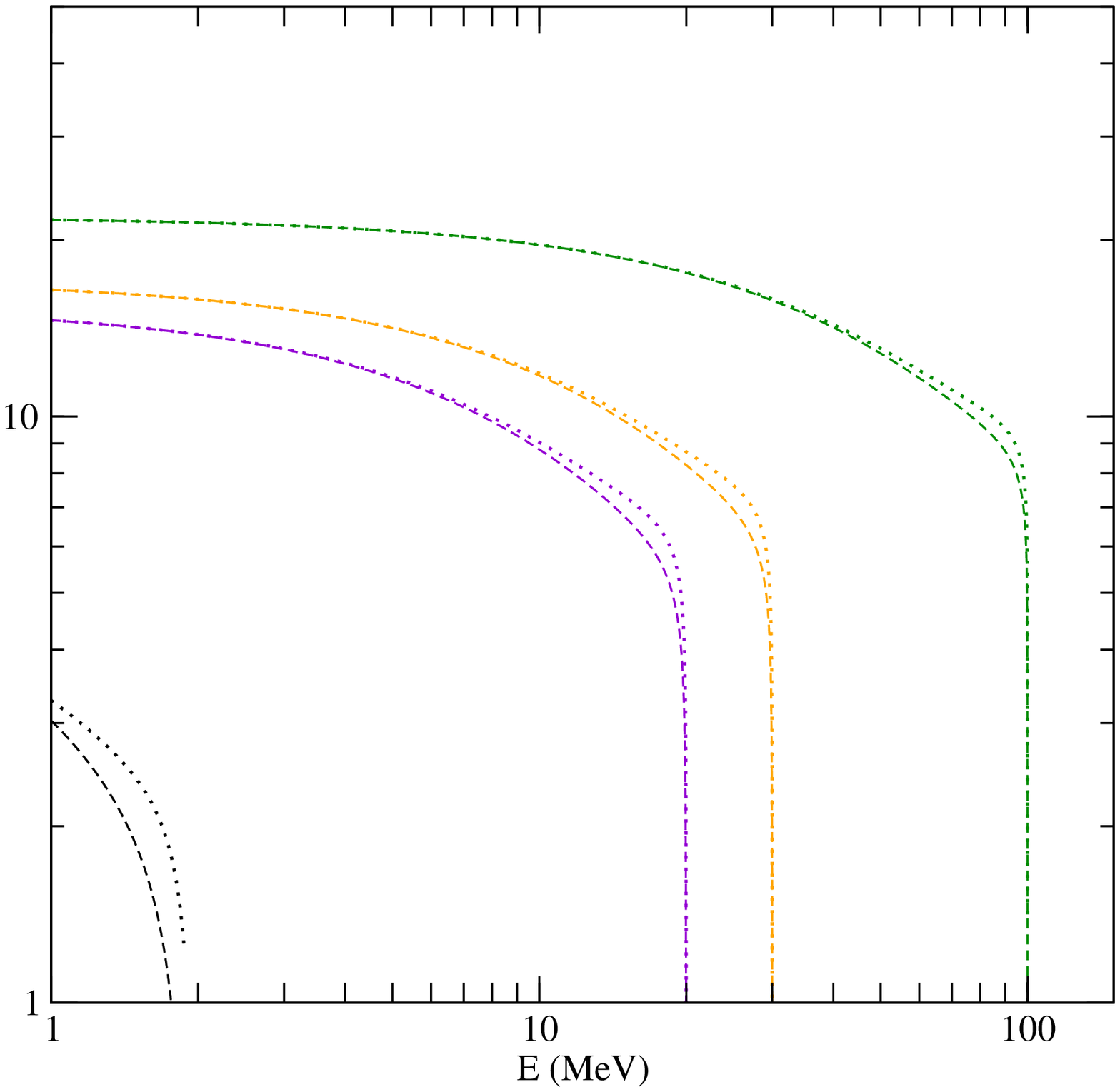} \\
 \includegraphics[angle=-90,width=11.cm]{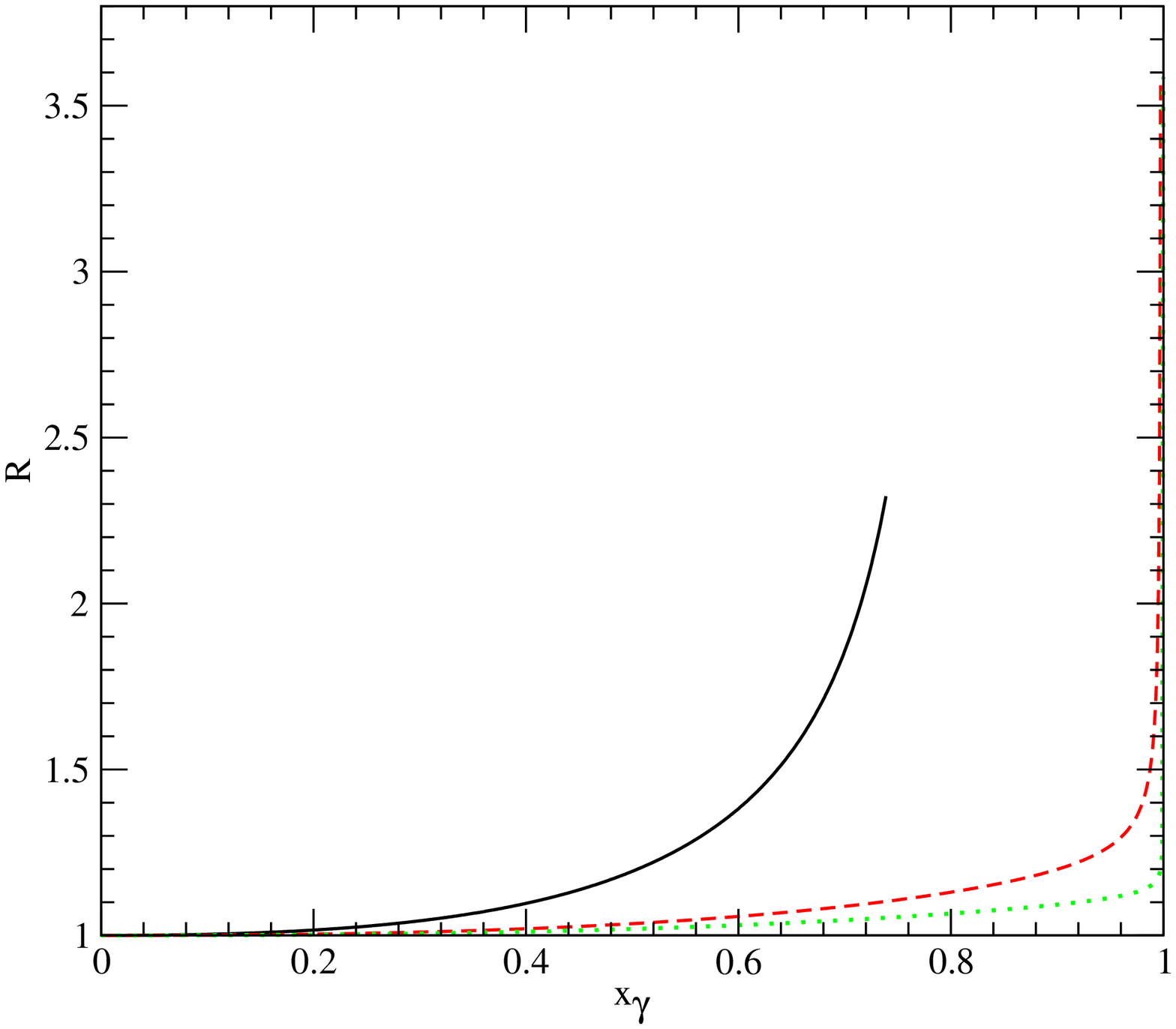}
    \caption{ Upper plot: our Bremsstrahlung cross section (dotted
lines) versus the Bremsstrahlung cross section used in Ref.~\cite{beacom05} (dashed lines) for 2, 20, 30 and 100
MeV. Lower plot: ratio of the two expressions  for 1, 10 and 100 MeV. }
    \label{ratioe}
    \end{center}
\end{figure}

In the upper plot in Figure~\ref{ratioe}, we show the Bremsstrahlung cross section as given in
Ref.~\cite{beacom05} (dashed lines) versus our expression (dotted lines) for 2, 20, 30 and  100 MeV. All our plots
are normalized to $\sigma_0  \, \frac{\alpha}{\pi} \, \frac{1}{\xg}$. To show more explicitly the difference
between these two expressions, we show in the lower plot the ratio:
$$ R =  \frac{\left(\frac{d \sigma}{d \xg}\right)_{\mbox{Eq.~\ref{eq:ScalarFSR}}}}{\left(\frac{d \sigma}{d \xg}\right)_{\mbox{Ref.~\cite{beacom05}}}}$$
for 1, 10 and 100 MeV.

The lower plot in Figure~\ref{ratioe} shows that the result obtained in this paper is up to a factor 3.6 larger
than the result obtained in Ref.~\cite{beacom05}. 
For maximal photon energy  ($\xg = 1-4z$), we obtain:
\begin{equation}
  \left.R\right|_{\xg=1-4z}
 = 1 + \frac{(1 - 4 \, z)^2}{(1+ 16 \ z^2) \
(\ln(4) - 1)}. \label{Req}
\end{equation}
 In the limit $z \to 0$,  Eq.~\ref{Req} tends to 3.6. Although the difference between
Ref.~\cite{beacom05} and the results presented in this letter becomes larger for large $\mdm$, the
phenomenological relevance might be larger for smaller $\mdm$. For large $\mdm$, the ratio is peaked at the
phase-space boundary in contrast with small $\mdm$ where a larger phase space region is affected.

\begin{figure}[htbp]
  \begin{center}
    \leavevmode
    \includegraphics[angle=-90,width=12.cm]{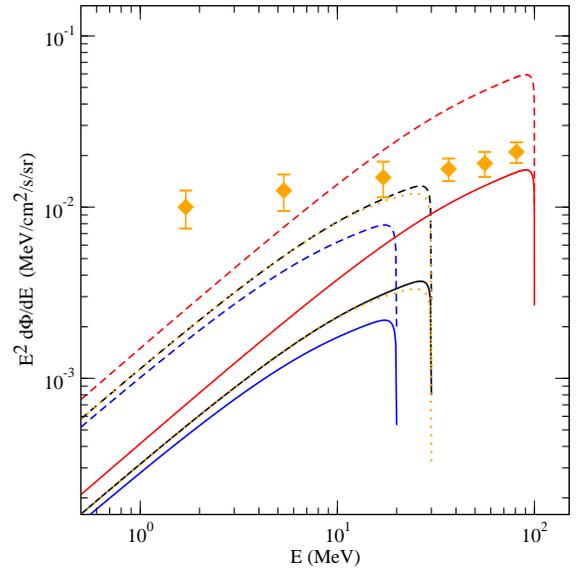}
    \caption{Bremsstrahlung flux for  20, 30 and 100 MeV, in the limit ($\mdm \ll \MF$), for $b < 5^\circ$ and $l < 5^\circ$ (dashed lines)
    and $b < 5^\circ$ and $l < 30^\circ$ (solid lines). The data are represented by
    diamonds and the result from Ref.~\cite{beacom05} is plotted for comparison (dotted lines).}
    \label{ratio}
  \end{center}
\end{figure}

The upper limit on the dark matter mass comes from the hard photon region. Fig.~\ref{ratio} shows the expected
flux of Bremsstrahlung photons for dark matter particles with a mass of 20, 30 and 100 MeV and  for $|b|< 5
^{\circ}$, $|l|<5^{\circ}$ (dashed lines) and $|b|< 5 ^{\circ}$, $|l|<30^{\circ}$ (solid lines) with $b$ the
latitude and $l$ the longitude. The expected flux in Ref.~\cite{beacom05} is represented by the dotted lines for
$\mdm = 30$ MeV.

Despite the sizeable difference in the hard photon limit the upper limit on the dark matter mass is not changed
significantly due to the limited precision of the data. We obtain the same result as in \cite{ascasibar}, i.e.
$\mdm < 30 \ \mbox{or} \ 100$ MeV, depending on whether one uses SPI data from regions defined by $|b|< 5
^{\circ}$ and $|l|<5^{\circ}$ (for which the 511 keV line intensity is about 0.018 ph/cm$^2$/s) or
$|l|<30^{\circ}$ (for which the intensity is 0.005 ph/cm$^2$/s) respectively.

Many papers, e.g. Refs.~\cite{komatsu,komatsuando,Rasera:2005sa} use the expression that is given in
\cite{beacom05}. It might be worth checking whether their results remain unchanged using the exact formula presented
in this work.

%%%%%%%%%%%%%%%%%%%%%%%%%%%%%%%%%%%%%%%%%%%%%%%%%%%%%%%%%%%%%%%%%%%%%%
\section{Conclusions}
%%%%%%%%%%%%%%%%%%%%%%%%%%%%%%%%%%%%%%%%%%%%%%%%%%%%%%%%%%%%%%%%%%%%%%
In light of Ref.~\cite{beacom05}, we computed the Bremsstrahlung emission
associated with the annihilation of spin-0 particle dark
matter particles and compared it with the result for $e^+ e^- \rightarrow \mu^+ \mu^- \gamma$.

We find that, in general, one cannot write the Bremsstrahlung cross section as the tree-level annihilation cross
section times a factor related to soft photon emission. In the limit where the mass of the new charged fermion $F$
(which is exchanged in the $t$-channel) is much heavier than the dark matter mass $\mdm$ and the electron mass
$\me$, we observe a factorization similar to that assumed by Beacom et al. However, the factor we find from the
explicit calculation is different from that given in Ref.~\cite{beacom05}. We thus conclude that the
bremsstrahlung cross section associated with the annihilation of
spin-0 particles into an electron--positron pair cannot be
obtained from final-state radiation in $e^+ + e^- \to \mu^+ + \mu^-$.

The difference between Ref.~\cite{beacom05} and our result is sizeable. However, the analysis presented in
Ref.~\cite{ascasibar} remains valid.

For $b < 5^\circ$ and $l < 5^\circ$, we find that the upper limit on
the dark matter mass is about 30 MeV. For $b < 5^\circ$ and $l <
30^\circ$, it is rather 100 MeV.

\section*{Acknowledgment}
We would like to thank W.~Bernreuther for useful discussions.

\section*{Appendix}
\def\Re{\mbox{Re}}
The squared amplitude for the Bremsstrahlung process is given by:
\begin{eqnarray}
  &&|M_\gamma|^2 = \frac{| M_{e^-}|^2}{D_{e^-}^2} +
  \frac{| M_{e^-}|^2}{D_{e^+}^2} + \frac{| M_{F} |^2}{D_{F}^2}
   \nn\\
  &+&
  2 \frac{\Re ( M_{e^-} M_{e^+}^{\ast}) }{D_{e^-} D_{e^+}}
  + 2 \frac{\Re( M_{F} M_{e^-}^{\ast})}{D_{F} D_{e^-}}
  + 2 \frac{\Re( M_{F} M_{e^+}^{\ast})}{D_{F} D_{e^+} },\nn\\
\end{eqnarray}
where
$M_i = e \epsilon^{\ast\mu } \ \bar{u}_e {\cal{M}}_i\, v_e  $
and
\begin{eqnarray}
  {\cal{M}}_{e^-} &=&
  \DiracMatrix{\mu} (\DiracSlash{\pg} + \DiracSlash{\pe} +\me)
  \left(\al \PL +\ar  \PR \right) \nn \\
  &\times&
  (\DiracSlash{\pe} + \DiracSlash{\pg} - \DiracSlash{\pdm} +\MF)
  \left(\al \PR + \ar  \PL \right),\\
  {\cal{M}}_{e^+} &=&
  \left(\al \PL + \ar  \PR \right)
  (\DiracSlash{\pe} - \DiracSlash{\pdm} + \MF) \nn \\
  &\times&  \left(\al \PR +  \ar  \PL \right)
  (-\DiracSlash{\pg} - \DiracSlash{\pp} + \me)
  \DiracMatrix{\mu}, \\
  {\cal{M}}_{F} &=&
  \left(\al \PL +  \ar  \PR \right)
  ( \DiracSlash{\pe} - \DiracSlash{\pdm} +  \MF)
  \nn \\
  &\times&
  \DiracMatrix{\mu} \
  (\DiracSlash{\pdms} - \DiracSlash{\pp} + \MF)
  \left(\al \PR + \ar  \PL \right),
\end{eqnarray}
with
\begin{eqnarray}
  D_{e^-} &=& \bigg((\pg-\pdm + \pe)^2 - \MF^2\bigg)
  \bigg(2 \pg\cdot\pe \bigg),\nn\\
  D_{e^+} &=& \bigg((\pe - \pdm)^2- \MF^2\bigg)
  \bigg(2\pg\cdot\pp\bigg), \nn\\
  D_{F} &=& \bigg((\pe - \pdm)^2- \MF^2\bigg)
  \bigg((\pp - \pdms)^2- \MF^2\bigg). \nn
\end{eqnarray}
\def\calA{{\cal A}}
\def\calB{{\cal B}}
\def\calC{{\cal C}}
\def\calD{{\cal D}}
\def\calE{{\cal E}}
\def\calN{{\cal N}}
\def\calR{{\cal R}}
\def\calV{{\cal V}}
\begin{eqnarray}
  \calA &=& -\me^4 \* \xg^2
  - 2 \* \mdm^4 \* (1 - x)\*(1 - \xb )\*(2 - 2\* \xg + \xg^2)
  \nn\\
  &+&  \mdm^2\* \me^2\* \big[4\* x\*\xb \* (1-\xg)
  -(2-\xg)\*(2-3\*\xg)  \big],
  \nn\\
  \calB &=& \me^6\*\xg^2
  - 2\* \mdm^2\* \me^4\* (1 - \xg)\*
  (2\*x\*\xb-2+2\*\xg+\xg^2) \nn\\
  &+& \mdm^4\* \me^2\* \big[-8 + 24\* \xg
  - 26\* \xg^2 + 9\* \xg^3 \nn\\
  &-& x\*\xb\* (-8 + 16\* \xg - 11\* \xg^2)
  \big] \nn\\
  &+& \mdm^6\* \big[-4\* x^4\* (1 - \xg)
  + 8\* x^3\* (2 - 3\* \xg + \xg^2)\nn\\
  &+&
  x^2\* (-20 + 36\* \xg - 18\* \xg^2 + \xg^3) \nn\\
  &+&  \xg\* (-4 + 10\* \xg - 9 \*\xg^2 + 3\* \xg^3) \nn\\
  &+& x\* (8 - 12\* \xg + 8 \*\xg^3 - 3\* \xg^4)\big],
  \nn\\
  \calC &=&
  \me^6\* \xg^2 \nn\\
  &-& 2 \* \mdm^2 \* \me^4\* (1 - \xg)\*
  \big[-2 - 2\* x^2 + 2\* x\* (2 - \xg) + 2\* \xg + \xg^2\big] \nn\\
  &+& 2\* \mdm^6\* (1 - x)\*(1 - \xg)\*
  \big[-2 + 4\* x^3 + 2 \*\xg + \xg^2 - \xg^3 \nn\\
  &+&  4\* x^2\* (-3 + 2\* \xg) +
  x\* (10 - 12\* \xg + 3\* \xg^2)\big]
  \nn\\
  &+& \mdm^4\* \me^2\*
  \big[-8 + 24\* \xg - 27\* \xg^2 +
  10\* \xg^3 \nn\\
  &-& 4\* x^2\* (2 - 4\* \xg + 3\* \xg^2)
  - 4\* x\* (-4 + 10\* \xg - 10\* \xg^2 + 3\* \xg^3)\big],
  \nn\\
  \calD &=&
  \me^6\* (1 - 2\* \xg)\* \xg^2 \nn\\
  &+& 2\* \mdm^6 \* (1 - x) \* (1 - 2\*\xg)^2
  \*(1 - x - \xg) \* (2 - 2 \*\xg + \xg^2) \nn\\
  &+&   2 \*\mdm^2 \*\me^4\*
  \big[2 - 8\* \xg + 8 \*\xg^2 - 2\* \xg^4 + x^2\* (2 - 6\* \xg + 3\*
  \xg^2)\nn\\
  &+& x\* (-4 + 14\* \xg - 12\* \xg^2 + 3\* \xg^3) \big]\nn\\
  &+& \mdm^4\* \me^2\* \big[-16
  - 8\* x^4\* (1 - \xg) + 64\* \xg - 97\* \xg^2 + 68\* \xg^3 \nn\\
  &-&18\* \xg^4
  + 16\* x^3 \* (2 - 3 \* \xg + \xg^2)
  + 2\* x^2\* (-28 + 64 \*\xg - 49\* \xg^2  \nn\\
  &+& 15 \* \xg^3) + 2 \* x\* (24 - 76\* \xg + 90\* \xg^2 - 51\* \xg^3
  + 11\* \xg^4)\big],
  \nn \\
  \calE &=& 2 \* \mdm^4 \*
  \big[( 1 - 2\* x)\* \mdm^2 + \me^2 - \MF^2 \big]^2 \* (1-x)^2 \nn\\
  &\times&\big[(1- 2\* \xb)\* \mdm^2 + \me^2 - \MF^2\big]^2 \*(1-\xb)^2,
  \nn \\
  \calN &=& -2\* \al^2 \* \ar^2 \* \me^2 \*
  \big[\me^2 + \mdm^2\* (1 - 2\*x)\big]
  \nn \\
  &\times&
  \big[\me^2 + \mdm^2\* (1-2\*\xb) \big] \* \calR +
  (\al^4 + \ar^4) \* \calV
  \nn \\
  \calR &=&
  \me^4\* \xg^2
  + 2\* \mdm^4\* (1 - x)\* (1 - \xb)\* \big[ 2 + \xg\* (-6 + 5 \*\xg) \big]
  \nn \\
  &+&
    \mdm^2\* \me^2
  \*\big[4 - 4\* x\* \xb \*(1 - \xg)
   + \xg\* (-8 + \xg\* (3 + 2\* \xg))\big],
  \nn \\
  \calV &=&
  -\me^{10}\* \xg^2
  - 2\* \mdm^8\* \me^2 \* (1 - x)\* (1-\xb)\* (1 - 2\* \xg)
  \nn \\
  &\times&
  \big[-6 + 8 \* x \*\xb\* (1 - \xg)
  + (10 - 3\* \xg) \* \xg\big] \nn\\
  &-&
   16\* \mdm^{10}\* (1 - x)^2 \* (1-\xb)^2 \* (1 - \xg)
   \* \big[(1 - x)^2 + (1 - \xb)^2 \big]
  \nn \\
  &+&
  \mdm^2\* \me^8\* \big[-4 + 4\* x\* \xb\* (1 - \xg)
  - \xg\* (-8 + \xg + 4\* \xg^2) \big] \nn\\
  &+& \mdm^4\* \me^6\*
  \bigg[12 - 2\* x\* \xb\*  \big(6 + \xg\* (-14 + 11\* \xg)\big)  \nn\\
  &-& \xg\* \bigg(40 + \xg\* \big(-47 + 2\* \xg\* (7 + 2\*
  \xg)\big)\bigg)
  \bigg] \nn \\
  &+& \mdm^6\* \me^4\* \bigg[4 - 16\* x^4\* (-1 + \xg) -
  32\* x^3\* (2 - \xg)\* (1 - \xg)
  \nn \\
  &+& \xg \* \bigg(8 + \xg \* (-51 + 8\* (8 - 3 \*\xg)\* \xg) \bigg)
    \nn \\
  &+&  4\* x^2\* \bigg(21 + \xg\*
  \big(-37 + \xg\* (14 + 3\* \xg) \big) \bigg) \nn\\
  &-& 4 \* x \* (2 - \xg)
  \*
  \bigg(5 + \xg\* (-5 + \xg \* (-6 + 7\* \xg) )\bigg)
  \bigg].
\end{eqnarray}

%\bibliography{FSR}

\end{document}